\documentclass{aastex}
\usepackage{natbib,epsfig,emulateapj5,rotate}

\tightenlines

\begin{document}

\newcommand{\psre}{\mbox{J1847$-$0130}}
\newcommand{\psr}{\mbox{J1847$-$0130 }}
\newcommand{\be}{\begin{eqnarray}}
\newcommand{\ee}{\end{eqnarray}}

\title{PSR~\psre: A Radio Pulsar with \\ Magnetar Spin Characteristics}
\author{M.\ A.\ McLaughlin\altaffilmark{1},  I.\ H.\ Stairs\altaffilmark{2},  V.\ M.\ Kaspi\altaffilmark{3}, D.\ R.\ Lorimer\altaffilmark{1}, \\ M.\ Kramer\altaffilmark{1},  A. G. Lyne\altaffilmark{1}, R.\ N.\ Manchester\altaffilmark{4}, F.\ Camilo\altaffilmark{5}, \\ G.\ Hobbs\altaffilmark{4}, A.\ Possenti\altaffilmark{6}, N.\ D'Amico\altaffilmark{7,8} \& A.\ J.\ Faulkner\altaffilmark{1}}
\altaffiltext{1}{Jodrell Bank Observatory, University of Manchester, Macclesfield, Cheshire, SK11 9DL, UK}
\altaffiltext{2}{Dept. of Physics and Astronomy, University of British
Columbia, 6224 Agricultural Road, Vancouver, BC V6T 1Z1 Canada}
\altaffiltext{3}{Department of Physics, McGill University, Rutherford Physics Building, 3600 University Street, Montreal QC H3A 2T8, Canada}
\altaffiltext{4}{Australia Telescope National Facility -- CSIRO,
P.O. Box 76, Epping NSW 1710, Australia}
\altaffiltext{5}{Columbia Astrophysics Laboratory, Columbia University, 550 W. 120th Street,
New York, NY 10027}
\altaffiltext{6}{Osservatorio Astronomico di Bologna, via Ranzani 1, 40127 Bologna, Italy}
\altaffiltext{7}{Universita degli Studi di Cagliari, Dipartimento di Fisica, S.P. Monserrato - Sestu Km 0,700, 09042 Monserrato (CA), Italy}
\altaffiltext{8}{INAF, Osservatorio Astronomico di Cagliari, Loc. Poggio dei Pini, Strada 54, 09012 Capoterra (CA), Italy}

\begin{abstract}
We report the discovery of PSR~\psre, a radio pulsar with a 6.7-s spin period,
in the Parkes
multibeam survey of the Galactic plane. The slowdown rate for the pulsar,
$1.3\times10^{-12}$ s s$^{-1}$, is high and implies a surface
dipole magnetic field strength of $9.4\times10^{13}$ G. This inferred dipolar
magnetic field strength is the highest by far among all known radio
pulsars and over twice the ``quantum critical field'' above which some
models predict radio emission should not occur. The inferred dipolar magnetic field strength
and period of this pulsar are in the same range as those of the
anomalous X-ray pulsars, which have been identified as being
``magnetars" whose luminous X-ray emission is powered by
their large magnetic fields. We have examined archival {\it ASCA} data
and place an upper limit on the X-ray luminosity of \psr which is lower than the
luminosities of all but one AXP. The
properties of this pulsar prove that inferred dipolar magnetic field strength and period
cannot alone be responsible for the unusual high-energy properties of the
magnetars and creates new challenges for understanding the possible relationship
between these two manifestations of young neutron stars.
\end{abstract}

\section{Introduction} \label{sec:intro}

According to the standard model, neutron stars
spin down according to
$\dot{\nu} \propto -\nu^{n}$, where $\nu$ is the rotation frequency, $\dot{\nu}$
is the frequency derivative
and $n$ is the ``braking index''. If the spin-down is due to  energy loss from
electromagnetic dipole radiation, then $n = 3$.
Magnetic field strengths at the neutron-star surface are then conventionally
inferred using the relation
\be 
B = 3.2\times10^{19} \sqrt{P\dot{P}} \hspace{0.1in} {\rm G}
\label{eq:bfield}
\ee
where $P = 1/\nu$ is the pulsar spin period, $\dot{P} = -\dot{\nu}/\nu^{2}$ and a
dipolar field structure, a neutron star radius of $10^{6}$~cm and moment of inertia 
$10^{45}$~g~cm$^{2}$ are assumed (e.g. Mancheser \& Taylor 1977). As a number of authors
have pointed out (e.g. Shapiro \& Teukolsky 1983; Usov \& Melrose 1995), this
relation gives the field strength at the magnetic equator, whereas the
perhaps more relevant field strength at the magnetic poles is a factor of
two higher. However, for consistency with earlier work, we use the
conventional value in this Letter. We note that no direct observational confirmation
has been made of inferred fields so they should be considered rough estimates only.
In particular, they are completely insensitive to higher order multipoles at the stellar
surface.

The spin parameters measured for neutron stars reveal inferred dipolar magnetic field strengths
 ranging from $10^8$~G for the ``recycled'' millisecond pulsars to $10^{15}$~G
for the anomalous X-ray pulsars (AXPs) and soft gamma-ray repeaters (SGRs). These so-called
``magnetars'' have spin periods ranging from 5.2~s to 11.8~s and inferred surface dipole magnetic
field strengths ranging from $5.9\times10^{13}$~G to $1.8\times10^{15}$~G. Their
X and gamma-ray luminosity is too great to be powered by the neutron star spin-down and
is attributed to the decay of their
ultra-strong magnetic fields \cite{td95,kou98,kou99,gkw02,kgw+03}.
Despite numerous
attempts, and aside from a radio detection of SGR~1900+14 by Shitov et al. (2000)
that has not been confirmed \cite{lx00},
these objects have not yet been been detected at radio
frequencies \cite{kbhc85,cjl94,llc98,gsg01}. This was explained by the
suppression of pair production above the quantum critical field 
$B_{cr}= m_e^2c^3/e\hbar \simeq 4.4\times 10^{13}$ G, at which the
cyclotron energy is equal to the electron rest-mass energy.  Above this
field, it was expected that pair production would cease due to ineffective
competition with the quantum electrodynamical process of magnetic photon splitting \cite{bh98}.

Prior to the Parkes Multibeam Pulsar Survey, the largest dipolar magnetic field strength
 inferred for any
radio pulsar was $2.1\times10^{13}$~G for PSR~B0154+61 \cite{antt94}.
The above paradigm held and a clear division in inferred dipolar
magnetic field strength between the radio pulsars and magnetars was
possible. 
This was challenged by the Parkes discoveries of PSRs J1744$-$3333
and J1814$-$1733 with implied dipole field strengths of $5.2\times10^{13}$~G and
$5.5\times10^{13}$~G \cite{mhl+02,ckl+00}.
In this Letter, we present the discovery of PSR~\psre, a radio pulsar with a
$9.4\times10^{13}$~G implied dipolar magnetic field strength.

\section{Radio Observations and Analysis} \label{sec:obsandresults}

PSR~\psr was discovered in the Parkes Multibeam Pulsar Survey of the Galactic
plane (e.g. Manchester et al. 2001), which has thus far discovered over 680 new pulsars.
The instrumental setup used for search and timing observations is described
in detail in Manchester et al.~(2001). The 6.7~s period of this pulsar is the
second-longest known; 
Fig.~1 shows the mean pulse profile. Following the calibration procedure described by
Manchester et al.~(2001), we calculate the flux
density at 1374 MHz to be 0.25 mJy, which for the distance of $\sim$ 8.4 kpc translates to a
radio luminosity of 18 mJy kpc$^{2}$. This distance is calculated from the dispersion measure and
a model for the Galactic electron density \cite{cl02}.
Because of systematic uncertainties and the possibility
of unmodeled interstellar medium features, both of which effects are greater for objects like
this pulsar which lies deep in the Galactic plane, we estimate that the distance may be
in error by as much as 30\%.

Timing observations since 2001 August 14 have resulted in the measurement of 22 times of arrivals.
 We fitted these using the TEMPO software
package\footnote{\tt http://pulsar.princeton.edu/tempo} to derive the timing model given in Table~1.
The post-fit timing residuals are featureless,
with no evidence for a binary companion, timing noise or period glitches.
The surface dipole magnetic field strength inferred from Eq.~\ref{eq:bfield} is
$9.4\times10^{13}$~G, the largest by far among known radio pulsars.
The pulsar is apparently young, with a characteristic age (assuming magnetic-dipole
spin-down) of 83~kyr. There is no cataloged supernova remnant associated
with this pulsar \cite{gre02}.

In Fig.~2, we plot the period and period derivative of
\psr along with those of radio pulsars, all five known AXPs and the two
SGRs with measured period derivatives. In Table~2, we give the periods and inferred dipolar 
magnetic field strengths of \psr
and the
AXPs. The similarity of the spin parameters of this pulsar
to the AXPs, in particular 1E~2259+586 and 4U~0142+61,
is obvious.

\section{X-Ray Observations and Analysis} \label{sec:xray}

Unlike all of the known AXPs, this
pulsar is not detected as a point source in the ROSAT All-Sky X-Ray Survey
\cite{vab+99}. In the best ROSAT data set obtained with the Position
Sensitive Proportional Counter instrument (sequence RP900402N00), the pulsar is 51$'$
 off-axis,
precluding a stringent upper limit calculation. However, a field containing the pulsar 
was observed with the
{\it ASCA} telescope \cite{tih94} on 1998 October 7 as part of a Galactic plane survey. 
In this 6~ks observation (Sequence Number 56000000),  
the pulsar position is 13$'$ from the pointing direction.
Given this offset, only the two Gas Imaging Spectrometer instruments (GIS2 and GIS3), 
which have a 25$'$ field of view,
are of use. Starting with the archived screened event files, we
first corrected for a known star tracker problem\footnote{Done using the
{\tt FTOOL offsetcoord} and coordinate offsets determined using a look-up
table available at {\tt
http://legacy.gsfc.nasa.gov/docs/asca/coord/updatecoord.html}.}. We then
considered a circular aperture of radius 4.5$'$ around the pulsar
position, as well as a neighboring region of the same area in order to
estimate the background, using the {\tt FTOOL xselect}.  No significant
emission above the background level was seen at the pulsar position.  We
use the method described by Pivovaroff et al. (2000)
to establish a 90\% confidence level upper limit on the
source count rate in the 2--10~keV band of $6 \times 10^{-3}$~counts s$^{-1}$.

In order to convert this upper limit on the count rate to a luminosity
upper limit, we assume isotropic emission and an equivalent neutral
hydrogen column density \cite{dl90} of $3 \times 10^{22}$~cm$^{-2}$.
All of the AXP spectra
can be fit by a model consisting
of a power-law and blackbody component, with spectral indices and blackbody
temperatures ranging from 2.4 $-$ 3.6 and 0.41 $-$ 0.65 keV, respectively
 \cite{gv97,tgsm02,pkw+02,pkw+03,ris+03}.
As shown in Table~2,
AXPs 1E~2259+586 and 4U~0142+61 have the spin-down parameters most similar to \psre.
The spectrum of 1E~2259+586 can be described by a power-law index of 3.6 and blackbody
temperature of 0.412~keV, with the power law contributing
roughly 50\% of the unabsorbed flux \cite{pkw+02}.
The spectrum of 4U~0142+61  can be described by a power-law index of 3.35 and blackbody
temperature of 0.458~keV, with the power law contributing
roughly 60\% of the unabsorbed flux \cite{pkw+03}.
Assuming either of these spectra and a distance of 8.4~kpc results in a 90\% confidence upper limit of
$\sim 5 \times 10^{33}$~ergs~s$^{-1}$
for the 2$-$10~keV unabsorbed luminosity $L_x$ of \psre. 

In Table~2\nocite{ggk+02}, this upper limit, taking into account the 30\% distance uncertainty, 
is compared with luminosities measured for the AXPs  
\cite{pkw+02,tgsm02,ggk+02,pkw+03,ris+03}, where AXP
distances from \"Ozel et al. (2001)\nocite{opk01} have been adopted.
The derived upper limit on the
2$-10$ keV luminosity of \psr is less than the luminosity of all of the AXPs
except 1E~1048$-$5937.
In Table~2, we also list the spin-down luminosities (i.e. $\dot{E} = 4\pi^{2} I\nu\dot{\nu}$, where $I$ is the moment of inertia), of \psr and the AXPs.
Becker \& Trumper (1997) find that, for radio pulsars which also are detected at X-ray
energies, $L_x/\dot{E} \sim 10^{-3}$. Our upper limit
for X-ray emission for \psr is still two orders of magnitude higher than the X-ray luminosity
expected given that relation.

\section{Discussion} \label{sec:discussion}

Despite the remarkable similarity
between the spin-down properties of \psr and the magnetars, their emission
properties are apparently very different. As discussed in \S1, searches
for radio pulsations from magnetars have so far been unsuccessful or
unconfirmed.  On the other hand, unlike all of the known AXPs, \psr
does not appear to be a strong X-ray source. However, because of the large uncertainties in 
calculating X-ray luminosities for \psr and the AXPs, X-ray observations of higher sensitivity
 are necessary
to determine if the luminosity does indeed lie well below those of all of the AXPs.

The discovery of \psr challenges those radio-pulsar emission models which
depend upon pair-production cascades above the magnetic poles and hence on
the strength of the magnetic field. This pulsar's inferred dipolar magnetic field lies more
than a factor of two above the quantum critical field above which some models predicted
the cessation of radio emission \cite{bh98}. As discussed in \S\ref{sec:intro}, 
magnetic field strengths calculated at the poles
 are a factor of two greater than the
conventionally quoted surface field value, exacerbating this discrepancy.
The radio luminosity at 1374~MHz of 18~mJy~kpc$^{2}$ is typical for observed radio
pulsars, indicating that photon splitting does not inhibit radio emission
at these magnetic field strengths.

A newer model \cite{zh00} argues that high-magnetic-field radio pulsars
and magnetars are all rotating high-field neutron stars but that their
magnetic axes have different orientations with respect to their rotation
axes.  This model predicts that radio pulsars should not have surface
dipole magnetic field strengths (inferred via Eq.~\ref{eq:bfield}) 
greater than $9.5\times10^{13}$ G, close to the
field of \psre.  However, this boundary is uncertain because it depends upon
roughly the square of the unknown neutron star temperature, with the model assuming a typical value of
$3\times10^{6}$ K.  Another possibility \cite{bh01} is that
only photons from one polarization split, allowing photons of the other
polarization to produce pairs.  Alternatively, the very different
properties of the magnetars and the high-field pulsars could be due to the
two populations having similar dipolar magnetic field configurations, but with
the magnetars also having quadrupole or other multipole components
\cite{gkw02}. Yet another possibility is that neutron stars spin down under the
combined effects of magnetic dipole spin-down and spin-down due to a propeller 
torque applied by a surrounding fallback disk, as suggested by Alpar et al. (2001).
This model predicts a small population of pulsars with long spin periods, high inferred
surface dipole magnetic fields and small characteristic ages. As these pulsars
would actually have conventional (i.e. $\sim 10^{12}$~G) magnetic fields and
 be very old, independent age estimates could distinguish between
propeller spin-down and pure magnetic dipole spin-down.

To understand the possible relationship between radio pulsars and magnetars and to
determine the physics responsible for their emission mechanisms, it is
essential to find more radio pulsars like \psre. While it is currently
the only known radio pulsar with inferred dipolar magnetic field strength and period as high as
those of the magnetars, there are significant survey selection effects
acting against the detection of long-period radio pulsars, which are
therefore likely to be under-represented in the known population. These
effects include high-pass filtering to increase the robustness of searches
to radio frequency interference and software cutoffs at long periods. This is
evidenced by our initial detection of \psr at half its true period and the
 detection of the 8.5-s pulsar J2124$-$3933 at a third of its period
\cite{ymj99}. Furthermore, the empirically determined relationship
between pulse width and pulse period \cite{ran93} implies that long-period pulsar
beams are narrower and hence less likely to sweep across the Earth. This is of course
somewhat balanced by the fact that narrower pulses have more detectable harmonics.
Additionally, assuming magnetic dipole spin-down, 
such pulsars may only be visible for the relatively short
time that they are young ($\sim 10^5$ yr) and in the magnetar-like region of
Fig.~2.

The discovery of this pulsar, with spin-down properties indistinguishable
from those of some AXPs, strengthens the conclusion of 
Pivovaroff et al. (2000) that the X-ray
emission properties of some AXPs must
depend on more than their inferred surface dipole magnetic field strengths. Similarly,
there is no reason based on period and period derivative
why magnetars cannot be radio emitters.  The
non-detections of radio emission should not be taken as proof that
magnetars are radio silent, as is evidenced by recent detections
\cite{cam03} of very faint pulsars with
luminosities of 0.5 mJy kpc$^{2} - 3$ mJy kpc$^{2}$, resulting from deep
targeted searches.  Alternatively, the magnetars may be strong radio emitters
whose radio beaming
fractions are small.
To determine whether or not \psr and the magnetars are
different manifestations of the same source population would be the
detection of radio emission from an AXP or SGR, or the detection of
magnetar-like high-energy emission from a high-field radio pulsar.

\acknowledgments

We thank Alice Harding for useful discussions. The Parkes Observatory is
part of the Australia Telescope which is funded by the Commonwealth of
Australia for operation as a National Facility managed by CSIRO. MAM is an
NSF-DRF postdoctoral fellow. IHS holds an NSERC UFA and is supported by a
Discovery Grant. VMK is a Canada Research Chair and is supported by
NSERC, NATEQ, CIAR and NASA. DRL a University Research Fellow supported by the Royal
Society. FC is supported by the NSF and NASA. 

{}

\begin{deluxetable}{lr}
\tablewidth{0pt}
\tablecaption{Measured and Derived Parameters of PSR~\psre}
\tablehead{Parameter & Value}
\startdata
Right ascension (J2000)   \dotfill & $\rm 18^{\rm h} 47^{\rm m}35\fs 18(6)$ \\
Declination (J2000)       \dotfill & $\rm -01^\circ  30'      44\farcs 0(16)$\\
Galactic longitude (degrees)  \dotfill & $31.15$ \\
Galactic latitude (degrees)  \dotfill & $+0.17$ \\
Barycentric Period, $P$ (s)  \dotfill &  6.7070459454(5) \\
Period derivative, $\dot{P}$ ($10^{-15}$ s s$^{-1}$)   \dotfill  & 1274.9(2) \\
Epoch (MJD)                \dotfill        & 52353\\
Timing data span (MJD)       \dotfill      & 52135--52571\\
Post-fit RMS timing residual (ms) \dotfill & 3.2\\
Dispersion Measure (cm$^{-3}$ pc) \dotfill & 668(7)\\
Flux Density at 1374 MHz, $S$ (mJy) \dotfill    & 0.25(4) \\
Width of pulse at 50\%, $W_{50}$ (ms) \dotfill	& 210 \\
\hline
Distance $d$ (kpc) \dotfill                    & $\sim$ 8.4 \\
Surface dipole magnetic field, $B$ ($10^{14}$ G) \dotfill &0.94\\
Characteristic age, $\tau_{c}$ (kyr) \dotfill           & 83\\
Spin-down luminosity, $\dot{E}$ (ergs s$^{-1}$) \dotfill & $1.7\times10^{32}$ \\
Radio luminosity at 1374 MHz, $Sd^2$ (mJy kpc$^2$) \dotfill         & 18\\
$2-10$ keV X-ray luminosity, $L_x$ (ergs s$^{-1}$) \dotfill               & $<5\times10^{33}$\\
\enddata
\label{tab:timingmodel}
\tablecomments{Figures in parentheses represent $1\,\sigma$
uncertainties in least-significant digits quoted. Distance derived from Cordes \& Lazio (2002).
 The characteristic age is calculated using the
standard magnetic dipole formula \cite{mt77} 
$\tau_{c}=P/2\dot{P}$. The spin-down luminosity $\dot{E} = 4\pi^{2} I\nu\dot{\nu}$, where the moment of inertia $I$ is assumed to be $10^{45}$ g cm$^{2}$.
}
\end{deluxetable}

\begin{deluxetable}{lccccc}
\tablewidth{0pt}
\tablecaption{Comparison with AXP Spin Parameters, Distances and X-Ray Luminosities}
\tablehead{Name    & $P$ (s) &  $B$ (10$^{14}$~G) &  $\dot{E}$ (ergs s$^{-1}$) & Distance (kpc)  & 2$-$10 keV $L_x$ (ergs s$^{-1}$)}
\startdata
1E 1048.1$-$5937 & 6.5 & 5.0 & $5.4\times10^{33}$ & $\ge$ 2.7 & $\ge$ $5\times10^{33}$ \\
1E 2259+586      &  7.0 & 0.59 & $5.6\times10^{31}$ & 4$-$7  & $4\times10^{34} - 1\times10^{35}$ \\
4U 0142+61       & 8.7 & 1.3 & $1.1\times10^{32}$ & $\ge 1.0$ or $\ge 2.7$  & $\ge 1\times10^{34}$ or $\ge 7\times10^{34}$  \\
RXS J1708$-$40   & 11.0 & 4.6 & $5.6\times10^{32}$ &  $\sim$ 8  &       $\sim   5\times10^{35}$ \\
1E 1841$-$045    & 11.8 & 7.1 &  $1.0\times10^{33}$ &   5.7$-$8.5       &       $2\times10^{34} - 5\times10^{34}$ \\
\hline
PSR~J1847$-$0130 & 6.7 & 0.94        & $1.7\times10^{32}$ &       $6-11$       & $< 3\times10^{33} - 8\times10^{33}$   \\
\enddata
\label{tab:compare}
\end{deluxetable}{}

\begin{figure}[ht]
\epsscale{0.5}
\plotone{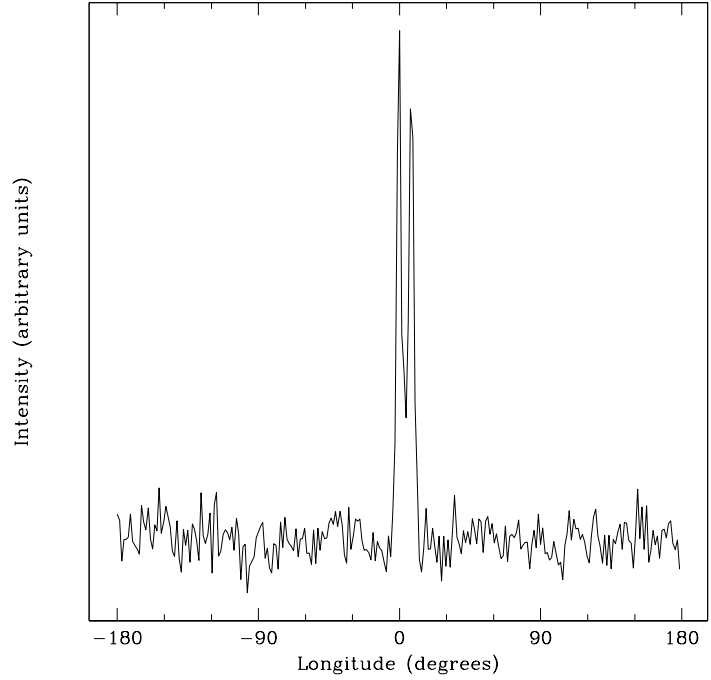}
\caption{
\label{fig:profile}
The integrated pulse profile of \psr at 1374~MHz. This profile was formed from roughly
2-hr of data taken over 11 epochs.  The profile
is double-peaked with a 3\% duty cycle, measured as the full width of the profile at half the maximum
pulse intensity.
}
\end{figure}

\begin{figure}[ht]
\epsscale{0.7}
\plotone{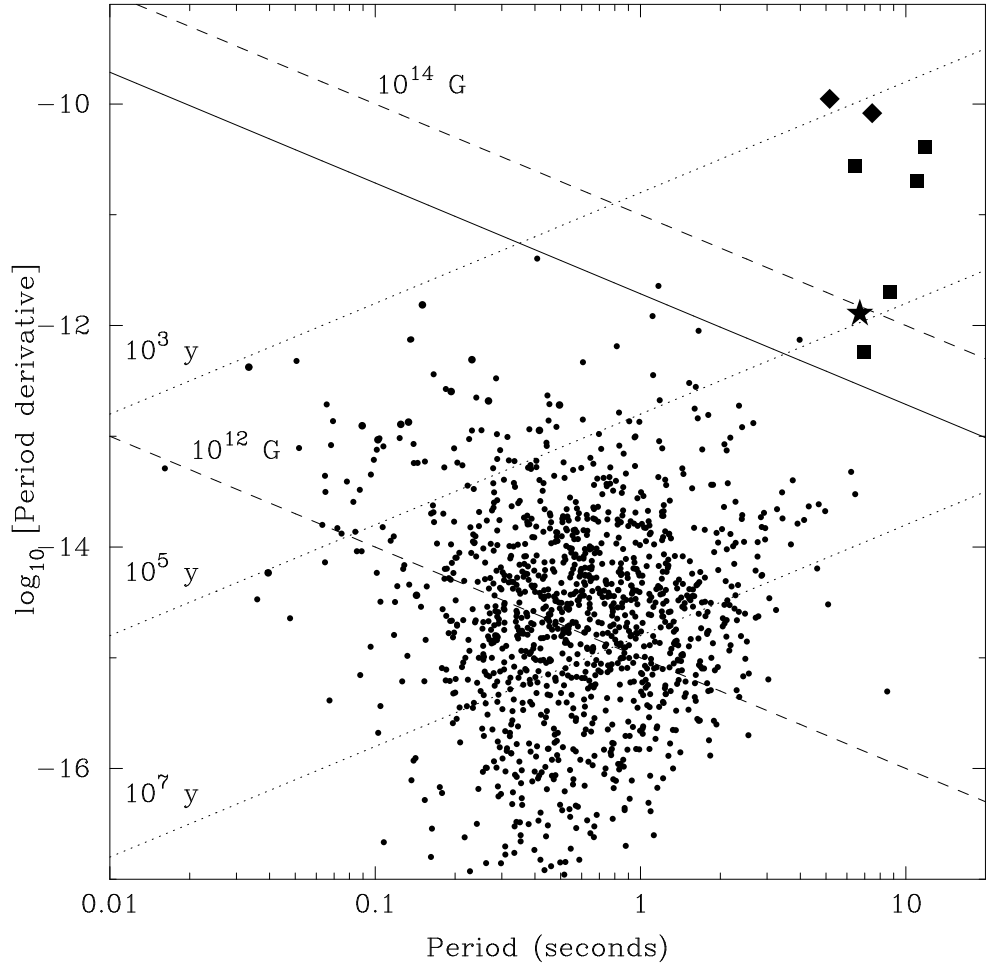}
\caption{
\label{fig:ppdot}
Period$-$period derivative for radio pulsars (dots), SGRs (diamonds) and
AXPs (squares). PSR~\psr is marked with a star. Pulsars with periods less than
10~ms and/or period derivatives less than $10^{-17}$ are not shown.
Lines of constant inferred dipolar magnetic field strength (dashed) and constant characteristic age
(dotted) are labeled. The solid line indicates the quantum critical field
of $4.4\times10^{13}$ G.
}
\end{figure}

\end{document}